\documentclass[prx,showkeys,showpacs,footnoteinbib,onecolumn,
unsortedaddress]{revtex4-2}
\usepackage{graphicx} 
\usepackage{dcolumn} 
\usepackage{amssymb}
\usepackage{amsmath}
\usepackage{physics}
\usepackage{natbib}
\usepackage{xcolor}
\usepackage{soul}
\usepackage{siunitx}
\usepackage{float}
\usepackage{esdiff}
\usepackage{bbold}
\usepackage{mathtools}
\usepackage{todonotes}
\usepackage{hyperref}
\usepackage{lineno}

\newcommand{\mytodo}[1]%
{{\todo[inline,backgroundcolor=blue!10!white]{#1}
}}

\colorlet{RED}{red}

\begin{document}

\title{Gate tunable edge magnetoplasmon resonators}

\author{Elric Frigerio$^{1}$}
\author{Giacomo Rebora$^{2}$}
\author{Mélanie Ruelle$^{1}$}
\author{Hubert Souquet-Basiège$^{2}$}
\author{Yong Jin$^{3}$}
\author{Ulf Gennser$^{3}$}
\author{Antonella Cavanna$^{3}$}
\author{Bernard Plaçais$^{1}$}
\author{Emmanuel Baudin$^{1}$}
\author{Jean-Marc Berroir$^{1}$}
\author{Inès Safi$^{4}$}
\author{Pascal Degiovanni$^{2}$}
\author{Gwendal F\`eve$^{1}$}
\author{Gerbold C. \surname{M\'enard}$^{1}$}
\email{email: gerbold.menard@phys.ens.fr}
\affiliation{$^{1}$ Laboratoire de Physique de l’Ecole normale sup\'erieure, ENS, Universit\'e PSL, CNRS, Sorbonne Universit\'e, Universit\'e Paris Cit\'e, F-75005 Paris, France}
\affiliation{$^{2}$ Univ Lyon, ENS de Lyon, Université Claude Bernard Lyon 1, CNRS, Laboratoire de Physique, F-69342 Lyon, France}
\affiliation{$^{3}$ Centre de Nanosciences et de Nanotechnologies (C2N), CNRS, Universit\'e Paris-Saclay, 91120 Palaiseau, France}
\affiliation{$^{4}$ Laboratoire de Physique des Solides-CNRS-UMR5802. University Paris-Saclay, B\^at.510, 91405 Orsay, France}
\begin{abstract}

Quantum Hall systems are platforms of choice to study topological properties of condensed matter systems and anyonic exchange statistics. In this work we have developed a tunable radiofrequency edge magnetoplasmonic resonator controlled by both the magnetic field and a set of electrostatic gates, meant to serve as a versatile platform for future interferometric devices designed to evidence non-abelian anyons. In our device, gates allow us to change both the size of the resonant cavity and the electronic density of the two-dimensional electron gas. We show that we can continuously control the frequency response of our resonator, making it possible to develop an edge magnetoplasmon interferometer. As we reach smaller sizes of our resonator, finite size effects caused by the measurement probes manifest. In the future, such device will be a valuable tool to investigate the properties of non-abelian anyons in the fractional quantum Hall regime.
\end{abstract}


\date{\today}

\maketitle

\section{Introduction}

Under strong perpendicular magnetic fields, two-dimensional electron gases (2DEGs) enter the quantum Hall regime. One important feature of this effect is the appearance of edge states with quantized conductance. Edge state transport can be described in terms of free collective bosonic  modes called edge magnetoplasmons (EMP) \cite{andrei1988low, volkov1988edge, safi1995, von1998, safi1999, senechal1999} that propagate along the edge with velocity $v$ \cite{ talyanskii1992spectroscopy, talyanskii1994experimental, kumada2011edge, petkovic2013carrier, kumada2014resonant, mahoney2017chip, martinez2023}.

When two counter propagating edges are brought together at the level of a quantum point contact (QPC), the granularity of charge carriers comes into play. In the regime of the integer quantum Hall effect (IQHE), the excitations tunneling at the QPC are quasi-electrons that obey the Fermi statistics. However, in the case of the fractional quantum Hall effect (FQHE), these excitations are no more fermionic but anyonic and follow an anyonic statistics.

Unlike fermions, anyons are quasiparticles with fractional charge and non-trivial exchange statistics. This means that the system acquires a non trivial braiding phase when two anyons are exchanged. Such properties were recently demonstrated experimentally in collider \cite{bartolomei2020fractional, ruelle2023, glidic2023} and interferometer \cite{nakamura2020direct, nakamura2023} experiments. Those experiments focused on the anyonic nature of the abelian, i.e. commutative, states $\nu=1/3$ and $\nu=2/5$. However, other fractional fillings are expected to host non-abelian anyonic excitations; in particular the $\nu=5/2$ filling factor \cite{stern2010}. The braiding of such non-abelian anyons has been proposed as a promising platform for topologically protected quantum computing protocols \cite{dassarma2006, nayak2008}.

One proposal to evidence these non-abelian properties is to study the absorption of microwave radiation by EMPs in an isolated Hall island \cite{cano2013microwave}. Because of its isolated nature, the Hall island is a resonant cavity for the EMPs. In this cavity, a resonance at $f=v/L$ (in the GHz range) is defined by the ratio between the velocity $v$ of the EMPs and the perimeter $L$ of the island \cite{talyanskii1992spectroscopy, talyanskii1994experimental, mahoney2017chip}. By adding a QPC, the resonator becomes a quasiparticle interferometer operated in the microwave domain. It is predicted that the statistics of the quasiparticles involved in the tunneling process determine the amplitude and period of the oscillations of the absorption with respect to the magnetic field. Cano {\it et al.} \cite{cano2013microwave} theorize that non-abelian properties, particularly expected to be observed at the $\nu=5/2$ filling, modulate this absorption depending on the parity of the number of quasiparticles present in both lobes of the interferometer. One of the main advantages of such a system is that the amplitude of the interferometric signal has a first-order dependence in the strength of the tunnel coupling. By contrast, other experiments relying on an open geometry of anyonic interferometers \cite{mcclure2009edge, mcclure2012fabry,nakamura2020direct, nakamura2023} make use of two QPCs and the interferometric signal only has a second order dependence on the tunnel coupling. Such a detection scheme thus increases the sensibility of the experiment and serve as an alternative route for the study of anyonic properties of quasiparticles in quantum Hall systems.

Aiming toward the realization of such an interferometer, we study here the microwave absorption of a bare Hall island, in a simple resonator regime, {\it i.e.} not in the interferometer mode. Contrary to previous works on EMP resonators \cite{andrei1988low, volkov1988edge, talyanskii1992spectroscopy, talyanskii1994experimental, kumada2011edge, petkovic2013carrier, kumada2014resonant, mahoney2017chip, martinez2023}, we define the cavity by a combination of electrostatic gates and chemical etching. This allows us to continuously change the shape of the cavity \cite{hashisaka2013}, making it possible to fully control the frequency of the resonant mode. We focus on micrometer-sized structures, approaching the dimensions required for interferometric experiments \cite{mcclure2009edge, mcclure2012fabry}, much smaller than previously considered EMP resonators \cite{talyanskii1992spectroscopy, hashisaka2013, mahoney2017chip}. In this configuration, we observe a deviation from the $f=v/L$ relation and we reach a regime in which the microwave measurement gates act as frequency filters of the resonant signal. We explore the finite size effect of the probes on the resonance frequency of the structure. We also consider the role of top gates on the propagation of magnetoplasmons and compare our results with a theoretical description based on an EMP scattering matrix formalism~\cite{safi1995, safi1999, bocquillon2013}.

\begin{figure}[h]
    \centering
    \includegraphics[width = 0.4\textwidth]{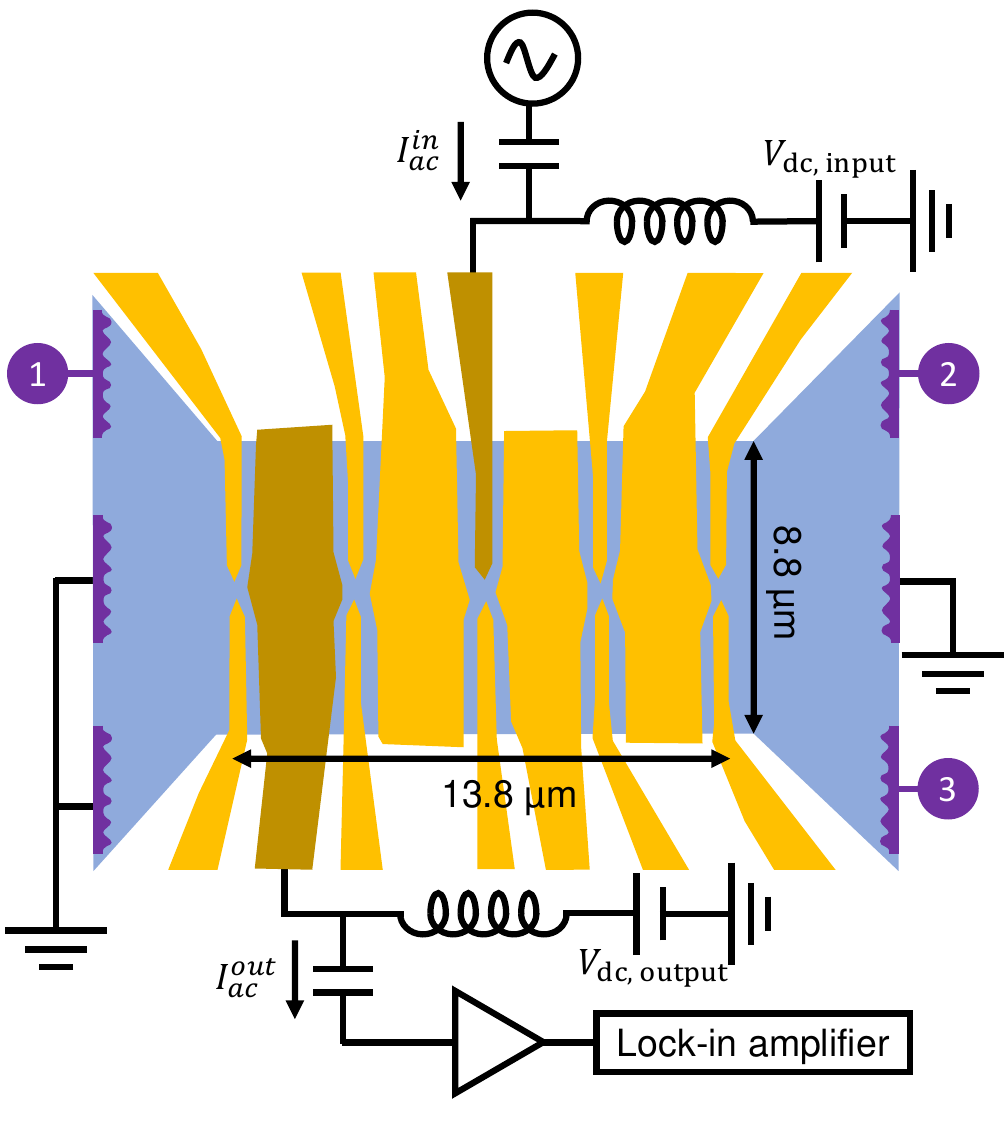}
    \caption{\textbf{Edge magnetoplasmon resonator:} Schematics of the sample. The 2DEG is shown in blue and the electrostatic gates are indicated in yellow. Two gates, shown in brown, are used both for electrostatic gating and microwave measurement. The injection of the microwave signal is done through the thin central brown gate and detection by the large brown pad on the left of the structure. Ohmic contacts (in purple) are also present on the sample far ($\approx 100~\mu m$) from the resonator area, in order to probe the dc properties of the sample (see supplementary note 2).}
    \label{fig1}
\end{figure}

\section{Results}
\begin{figure*}
    \centering
    \includegraphics[width = 0.8\textwidth]{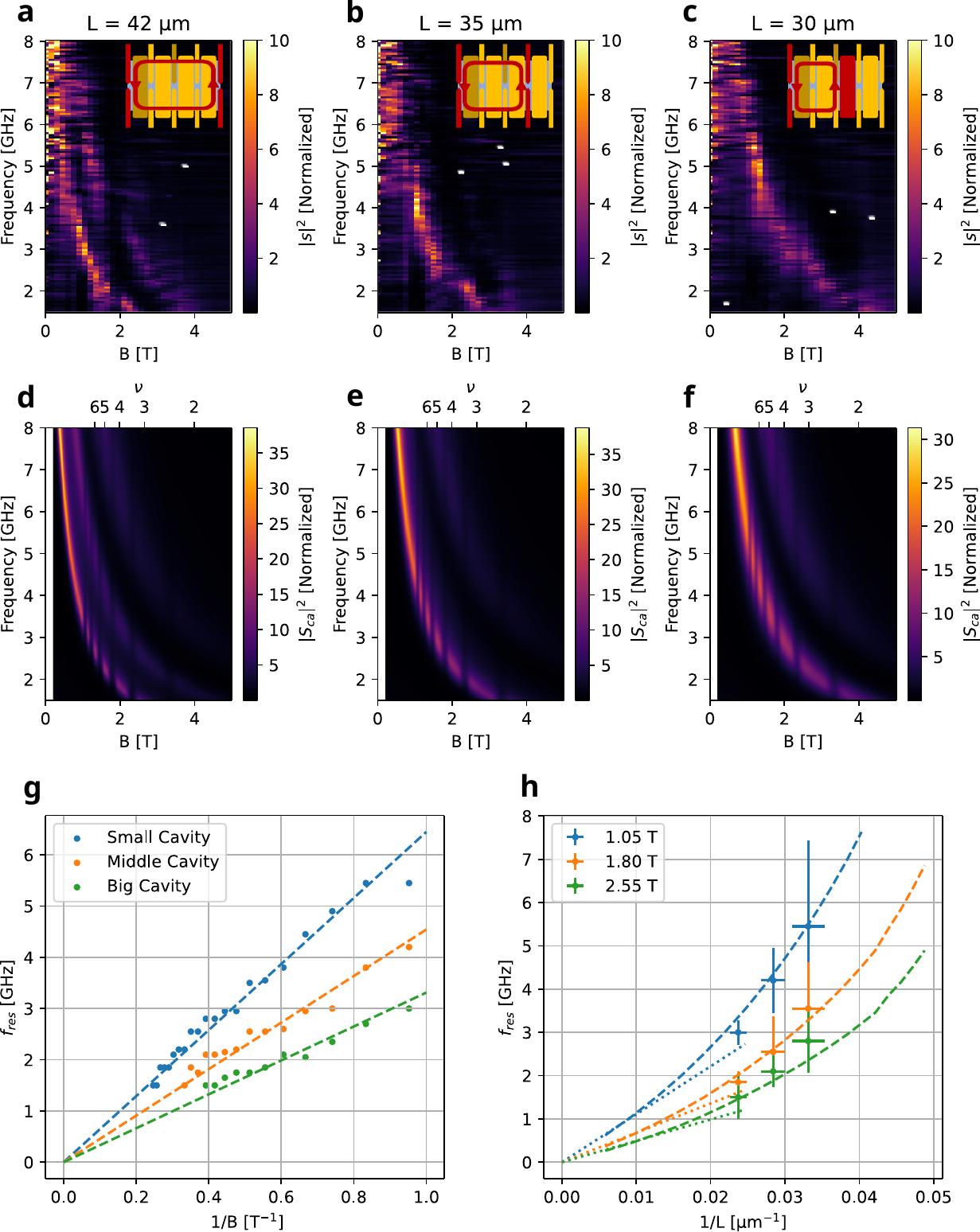}
    \caption{\textbf{Effect of the magnetic field and the size of the cavity:} (a)-(c) Color maps of the microwave transmission $|s(B,f)|^2$ of the sample as a function of the magnetic field and the excitation frequency for three different electrostatic configurations. From left to right the size of the cavity decreases (the electrostatic configuration is schematically represented in the inset of the figures) and, reciprocally, the resonance frequency increases. (d)-(f) Numerical simulations performed with the same cavity sizes as in (a)-(c). The top horizontal axis indicates the values of the filling factors associated to the density and field used in the simulations. (g) Frequency of the resonant mode as a function of the inverse of the magnetic field for the three cavities presented in (a) to (c). These points were obtained by extracting the maximum of the transmission signal at each value of the magnetic field. The dashed lines correspond to the linear fit of the experimental points. (h) Frequency of the resonant mode as a function of the inverse of the perimeter of the cavity for three different magnetic fields. The experimental data is presented with error bars and the dashed lines correspond to the result of the simulation. The dotted lines indicate the expected $f = v\left(B\right)/L$ behavior, only valid for large cavities, using for $v$ the values used to plot figures d-f. The error bars shown in this graph correspond, in the frequency axis, to the fitted width of the resonance. Those in the $1/L$ axis were taken by estimating the uncertainty on $L$ and propagating the error to $1/L$.}
    \label{fig2}
\end{figure*}

\textbf{Sample description.}
The Hall island considered here is based on an AlGaAs/GaAs heterostructure hosting a high mobility 2DEG. Ti/Pt/Ti/Au gates are evaporated on this structure. These gates are coupled only capacitively to the 2DEG (see figure \ref{fig1}). This geometry allows us to realize islands of controllable sizes, from $8\times\SI{7}{\micro\meter^2}$ ($L=\SI{30}{\micro\meter}$) to $8\times \SI{13}{\micro\meter^2}$ ($L=\SI{42}{\micro\meter}$), by choosing which QPC to open or to close (see insets of figures \ref{fig2}, a-c).

The injection and detection of the microwave signal is done through two bias-tees connected respectively to the top gate of the central QPC and to the left pad gate (in brown in figure \ref{fig1}). These two gates can also be addressed in dc and thus allow us to tune the perimeter of the structure as well as the local electronic density inside the resonator. Ohmic contacts are also present (shown in purple in figure \ref{fig1}). Their purpose is to control the opening and closing of the gates in the dc regime in the calibration stage of the experiment.

The large gates deposited on top of the structure have two roles: first, they allow us to tune the density of the 2DEG and second, they screen the inter-channel coupling, leading to a decrease of the velocity $v$ of EMPs by approximately one order of magnitude \cite{zhitenev1994, talyanskii1992spectroscopy, johnson2003}. Reducing the velocity brings the resonance frequency $f_\text{res}$ into our experimental frequency window of $0.5-\SI{8.5}{\giga\hertz}$ for magnetic fields between $0$ and $\SI{8}{\tesla}$.

\textbf{Discrete control of the cavity size.} Figures~\ref{fig2}.a.-c. present the result of the measurement of the microwave signal $s(f,B)$ transmitted through the cavity in three different electrostatic configurations of the same sample. The quantity plotted in this figure is the square of the normalized transmitted intensity $|s|^2$ (see methods section for details about the normalization procedure). In this figure, the frequency varies from $1.5$ to \SI{8}{\giga\hertz} and the magnetic field is swept between $0$ and \SI{5}{\tesla}, corresponding to filling factors $\nu \geq 2$. Each color map corresponds to the electrostatic configuration drawn in the inset of the figures where the QPCs used to define the cavity are represented in red. The main feature of these transmission maps (Figs.~\ref{fig2}.a-c) are the resonances whose frequency decreases with the magnetic field, following the magnetic field behavior of $v\propto 1/B$ \cite{talyanskii1990edge}. In Fig.~\ref{fig2}.g, the resonance frequency is plotted as a function of $1/B$ for three different cavity sizes (perimeters of $\SI{30}{\micro\meter}$, $\SI{35}{\micro\meter}$ and $\SI{42}{\micro\meter}$). The extrapolation of the linear fits down to $\SI{0}{\tesla^{-1}}$ demonstrates the inverse proportionality between the frequency and the magnetic field.

The full maps of the transmitted signal (Fig.~\ref{fig2}.a-c) exhibit discontinuities in the line shape following the resonance. These discontinuities are caused by the transitions between various plateaus of the integer quantum Hall effect \cite{mahoney2017chip}. This is particularly visible around $\SI{2}{\tesla}$. Within a plateau, the edge states are well defined thanks to the conductance suppression through the bulk of the resonator. However, between two plateaus, the bulk becomes conducting and the edge states cease to exist as 1D chiral conducting regions. Consequently, there is no well defined resonance frequency anymore and this manifests as discontinuities in our experimental maps. Figures.~\ref{fig2}.d.-f. present the corresponding numerical simulations that will be detailed below.

In the largest cavity ($L=\SI{42}{\micro\meter}$), we observe a second mode present at twice the frequency of the fundamental. This mode is not observed for smaller cavities. As we will show with our theoretical description, this is due to the finite-size of the detection gate that leads to ``blind spots'' in the EMP detection. When decreasing the size of the cavity, the frequency of the resonance increases and so does that of the first harmonic. The harmonic is then pushed into the blind spot of the detection and thus disappears in figures \ref{fig2} b and c.

For a fixed magnetic field $B$, reducing the perimeter of the cavity increases the resonance frequency. This is better seen by extracting the frequency at a given field and plotting it as a function of the inverse of the total perimeter $L$ of the cavity under study as shown on Fig.~\ref{fig2}.h for the fields $\SI{1.05}{\tesla}$, $\SI{1.8}{\tesla}$ and $\SI{2.55}{\tesla}$. If the frequency were proportional to the inverse of the total length of the cavity, we would observe a proportionality relation between $f_{\text{res}}$ and $1/L$ (dotted lines). However, linear fits to the data do not converge to 0. This peculiar behavior is caused by finite size effects of the input and output gates that will be explored theoretically in the next sections.

\begin{figure}
    \centering
    \includegraphics[width = 0.45\textwidth]{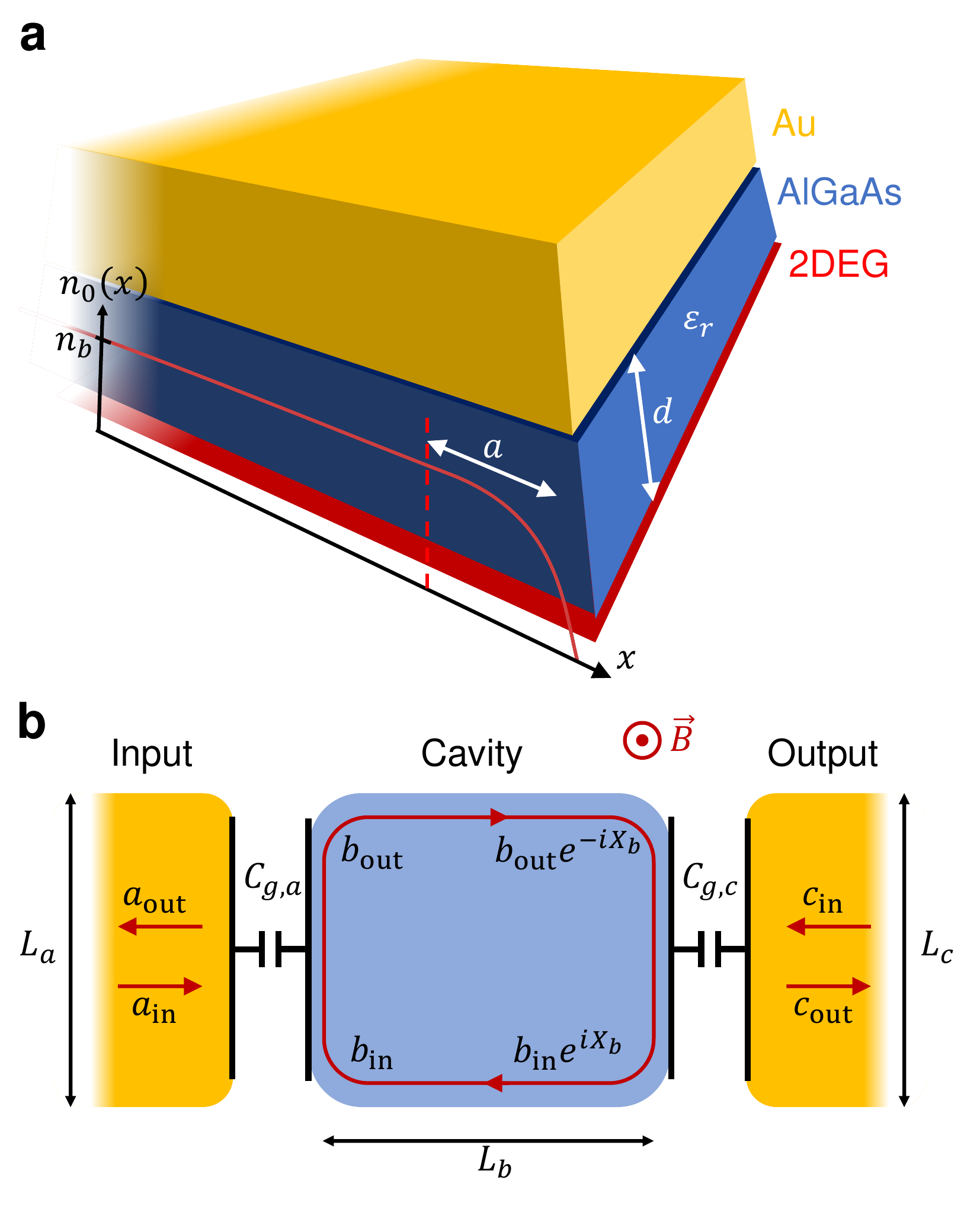}
    \caption{\textbf{Theoretical description of the system:} (a) The Au gate (in yellow) is positioned at a distance $d\approx \SI{105}{\nano\meter}$ above the 2DEG (in red) separated by the AlGaAs layer (in blue) with relative dielectric permittivity $\varepsilon_r$. The electronic density $n_0(x)$ is constant in the bulk and decreases to zero on a scale $a$ on the edge. (b) The radiofrequency transport through the system is described in terms of three different sections. The input (resp. output) gate is described by the incident and reflected amplitudes $a_\text{in}$ and $a_\text{out}$ (resp. $c_\text{in}$ and $c_\text{out}$). The cavity is connected capacitively to these two input and output gates of respective lengths $L_a$ and $L_c$ through capacitances $C_{g,a}$ and $C_{g,c}$. The magnetoplasmons acquire an amplitude $\mathrm{e}^{\mathrm{i}X_b}$ where $X_b=k(\omega)L_b$ when propagating along the edge of the cavity on a length $L_b$. The transmission amplitude is given by the term $S_{ac}$ in the magnetoplasmon scattering matrix. For readability reasons, the top gates are not represented here. Their effect is already taken into account by Eq.~\ref{eq/k-formula}.}
    \label{fig3}
\end{figure}

\textbf{Propagation of the EMPs.} The theoretical description of the system is done in two steps. First, we consider the propagation of magnetoplasmons in the presence of a gate following the description of Ref.~\cite{johnson2003}. We consider a 2DEG separated from the gate by a distance $d$ (see Fig.~\ref{fig3}.a). In between the 2DEG and the gate, the AlGaAs layer is described by a dielectric of relative permittivity $\varepsilon_r$ approximately equal to $12.4$ at low temperature \cite{strzalkowski1976, moore1996infrared, krupka2008}. The density $n_0(x)$ on the edge of the system vanishes over a length $a$ where it connects to the vacuum. 
%

The transmission amplitude for the charge edge magnetoplasmon at frequency $\omega/2\pi$ propagating over a distance $l$ is of the form $e^{ik(\omega)l}$. The real part of $k(\omega)$ describes the accumulated phase or equivalently the charge EMP effective velocity ($\Re(k(\omega))=\omega/v$), whereas its imaginary part $\Im(k(\omega))\geq 0$ describes the charge magnetoplasmon attenuation along propagation. 

In Ref.~\cite[Sec. IV.A.2]{johnson2003}, dissipation is associated with the dc longitudinal resistance, which is non-vanishing between the plateaus of the quantum Hall effect. However, this is not the only source of dissipation in quantum Hall systems. Dissipation also arises from the capacitive coupling to charge puddles within the bulk of the 2DEG, thereby taking energy away from the charge edge magnetoplasmon mode and ultimately dissipating it into the phonon bath \cite{kumada2014resonant,lin2021}. Such dissipation is present even when the longitudinal dc resistance vanishes within quantum Hall plateaus. At the lowest order, a contribution to $\Im(k(\omega))$, proportional to $\omega^2$ can be added to describe the damping of EMP along their propagation \cite{hashisaka2013}. Adding this capacitive dissipation to the expressions of 
Ref.~\cite[Sec. IV.A.2]{johnson2003} for the complex valued $k(\omega)$ for a wide edge leads to:
\begin{subequations}
\begin{align}
\label{eq/k-formula}
    k(\omega)&=\frac{\omega}{v}+i\left(\frac{r^2}{4\omega_c\tau a}+\frac{\xi\omega^2}{v^2}\right)\\
    v&=\frac{\gamma\omega_c a}{1+\gamma(r^2/4)+(1/\omega_c^2\tau^2)}
   \label{eq/velocity}
\end{align}
\end{subequations}
where $r\simeq 2.405$ is the first zero of the Bessel function $J_0$, $\omega_{c}=eB/m$ is the cyclotron pulsation and $\gamma = l_0 d/a^2$ with the characteristic length $l_0 = e^2 n_b/\varepsilon m\omega_c^2$. In our experiment, $n_b$ denotes the bulk electronic density and $\varepsilon = \varepsilon_r\varepsilon_0$ with $\varepsilon_0$ the dielectric permittivity of vacuum and $\epsilon_r = 12.4$ the relative dielectric permittivity of GaAs \cite{levinshtein99}. The effective mass $m$ of electrons in the 2DEG is taken to be $0.067$ times the mass of free electrons \cite{coleridge1996, fu2017}. $d$ is the distance from the top gates to the 2DEG,  $d = $~\SI{105}{\nano\meter}.  As we will discuss later, we obtain a best fit of our data by using $a=$\SI{2.8}{\micro\meter}. It leads to $\gamma = 1.4 \times 10^{-2} $ with $l_0 =$~\SI{1.1}{\micro\meter} at \SI{1}{\tesla} and makes equations \ref{eq/k-formula} and \ref{eq/velocity}  valid since  $\gamma \ll 1 $ and $d\ll a$. We reach a maximal value of $ka\simeq 1$ for the smallest cavities which should limit the relevance of some approximation made in \cite{johnson2003} and therefore impact our calculation. However, as we will see the final agreement with our data remains good.

The relaxation time of the Drude model $\tau$ is related to the dc longitudinal resistance and thus depends on the magnetic field (see supplementary note 5). Finally, the length $\xi$, which has to be fitted from the experimental data, accounts for dissipation through capacitive couplings via the $\xi \omega^2/v^2$ contribution to $\Im(k(\omega))$ where $v$, given by Eq.~\ref{eq/velocity} is the effective velocity of the magnetoplasmonic charge mode.

\textbf{Transmission model.}
Assuming that dissipative propagation of the charge EMP is described according to the previous paragraph, we can now turn our attention to the coupling of the resonator to the probing gates. These two gates are described as two microwave transmission lines, capacitively coupled to the 2DEG. The microwave propagation along these lines is treated within the input/output formalism for the setup depicted on Fig~\ref{fig3}.b.

This corresponds to a plasmonic Fabry-Perot (FP) cavity with extended mirrors corresponding to the region where its resonant modes couple to the external transmission lines through a capacitive coupling $C_{g,a}$ at the input gate of length $L_a$ and $C_{g,c}$ to the output gate of length $L_c$. Propagation along the uncoupled parts of the island over a distance $L_b$ leads to an amplitude $\mathrm{e}^{\mathrm{i}X_b}$ with $X_b=k(\omega)L_b$ (dissipation being contained in $\Im(k(\omega))$).

Microwave propagation across the FP cavity can then be described using an EMP scattering matrix connecting the incoming to the outgoing modes
\begin{equation}
    \left(\begin{matrix}
        c_{\text{out}}\\
        a_{\text{out}}
    \end{matrix}\right)
    =\left(\begin{matrix}
        S_{cc} & S_{ca}\\
        S_{ac} & S_{aa}
    \end{matrix}\right)\left(
    \begin{matrix}
        c_{\text{in}}\\
        a_{\text{in}}
    \end{matrix}\right)
\end{equation}
involving four scattering amplitudes whose explicit forms are given in supplementary note 6. The resonant mode of interest is accessible through the full transmission amplitude connecting $a_\text{in}$ to $c_\text{out}$:
\begin{equation}
S_{ca}=\frac{t'_a t_c\, \mathrm{e}^{\mathrm{i}X_b}}{1-r'_a r'_c \,\mathrm{e}^{2\mathrm{i}X_b}}.
\label{transampl}
\end{equation}
In this equation, $t_j$ and $r_j$ respectively denote the EMP transmission and reflection coefficients at the input ($j=a$) and output ($j=c$) mirrors of the FP cavity. For simplicity, these expressions assume that the distance traveled by the magnetoplasmons between the input and output gates and vice-versa are the same, as depicted on Fig.~\ref{fig3}.b. For an asymmetric sample and with the propagation amplitude discussed in the previous paragraph, the expressions are almost the same: first of all, the product of the two amplitudes in the denominator is $\mathrm{e}^{2\mathrm{i}X_b}$ where $X_b$ is computed from $L_b$ being the average of the propagation distances from $a$ to $c$ and $c$ to $a$. However, the exponential at the numerator has to be replaced by $\mathrm{e}^{\mathrm{i}k(\omega)L_{ac}}$ where $L_{ac}$ is the actual length of the path from $a$ to $c$.


To obtain the expressions of the transmission and reflection coefficients, we model the mirrors of the FP by a capacitive coupling between a transmission line with characteristic impedance $Z=$\SI{50}{\ohm} and the edge channel of the Hall island~\cite{feve2008}. We introduce dissipation for the charge edge magnetoplasmon mode beneath the metallic input and output gates as discussed in the previous paragraph. We assume total electrostatic mutual influence between the two gates and the resonator within each mirror of the FP cavity. This hypothesis is reasonable for our sample and ensures maximal coupling between the resonator and its left and right ports. As shown in supplementary note 6, this leads to the plasmonic transmission and reflection amplitudes:
\begin{align}
    t'_j&=- \mathrm{Re}(k(\omega)) L_j \sqrt{\mathcal{R}}\,\mathrm{sinc}\bigg(k(\omega)\frac{L_j}{2}\bigg)\,D_j^{-1} 
    \label{trans_coeff}, \\
     t_j&=- k(\omega) L_j \sqrt{\mathcal{R}}\,\mathrm{sinc}\bigg(k(\omega)\frac{L_j}{2}\bigg)\,D_j^{-1},
         \label{trans_coeff_2} \\
    r'_j&=\mathrm{e}^{-\mathrm{Im}(k(\omega))\frac{L_j}{2}}\bigg[1 +\mathrm{e}^{-\mathrm{i} k(\omega)\frac{L_j}{2}}\times \nonumber \\
&\times\alpha_j\, \mathrm{sinc}\bigg(k(\omega)\frac{L_j}{2}\bigg)\,\bigg(1-\mathrm{Re}(k(\omega))\frac{\mathcal{R}}{2}\frac{L_j}{\alpha_j}\bigg)\bigg]D_j^{-1},
    \label{refl_coeff}
\end{align}
with $D_j$ being the denominator coefficient
\begin{equation*}
\begin{split}
    D_j&=\mathrm{e}^{\mathrm{Im}(k(\omega))\frac{L_j}{2}}\bigg[1+\mathrm{e}^{\mathrm{i}k(\omega)\frac{L_j}{2}}\times \\ &\times \alpha_j\, \mathrm{sinc}\left(k(\omega)\frac{L_j}{2}\right) \left(1-\mathrm{Re}(k(\omega))\frac{\mathcal{R}}{2}\frac{L_j}{\alpha_j}\right)\bigg],
\end{split}
\end{equation*}
where $\mathcal{R}=2e^2Z/h$. The dimensionless coupling
\begin{equation}
    \alpha_j = \frac{e^2}{h}\frac{L_j}{v\, C_{g,j}}
\end{equation}
depends on the magnetic field via the EMP velocity $v$ and characterizes the importance of Coulomb interactions within the mirrors. 

Note that dissipative propagation of the charge EMP introduces a non-reciprocity in the scattering matrix between the transmission line and the edge channel (i.e. $t'_j\neq t_j$). This comes from the fact that when the transmission line is driven, it imposes a uniform potential all along the edge channel beneath the gate whereas, when an ac current is injected into the edge channel, the associated charge density wave is damped while propagating beneath the metallic gate and therefore induces a smaller voltage drop at the capacitor and thus a smaller drive on the transmission line. The non-reciprocity is a consequence of dissipation of the charge EMP along its propagation and may be used to probe it.

Now equipped with this theoretical description, we come back to the experimental data in which $|s|^{2}$ can be compared to $|S_{ca}|^2$. Figs.~\ref{fig2}.d-f depict the result of the calculation for three cavities with sizes equal to those of the cavities studied experimentally. In order to better compare the amplitude of the signal, we apply the same normalization procedure to the numerical simulations and to the experimental data (see methods section). We observe that we reproduce well the experimental data by considering a lineic capacitive coupling $c_{g}= \SI{3}{\nano\farad\per\meter}$ ($C_{g,j}=c_{g}L_{j}$). 
The electronic density is fixed by the dc transport data that indicates the position of the plateaus and leads to $n_b = \SI{1.93e11}{\centi\meter^{-2}}$ (see supplementary note 2). Finally, the best fit to the resonance frequency was obtained by using the value $a=\SI{2.8}{\micro\meter} \gg d$ for the length over which the electronic density falls to zero on the edge of the sample. This value is compatible with the ones obtained in Refs.~\cite{zhitenev1994, gourmelon2023}. The value of the dimensionless coupling $\alpha_j$ at $B= \SI{1}{\tesla}$ is $\alpha_j\approx 0.12$. It should be noted here that, in order to obtain a good agreement between data and theory, we consider edge modes propagating $\SI{400}{\nano\meter}$ inwards compared to the lithographically defined edges. This condition can be justified by the fact that we are in the wide edge limit given the value of $a$: the EMP modes emerge from the topological edge channels that are spatially separated from each other and are not found on the exact lithographic edge of the structure \cite{armagnat2020}. Using these parameters, we obtain velocities varying from $1\times10^{5}$~\SI{}{\meter\per\second} at \SI{1}{\tesla} to $2.1\times 10^4$~\SI{}{\meter\per\second} at \SI{5}{\tesla} (see appendix D) and wavevectors whose maximum value in our experimental range is  $2.4 \times 10^{6} $ m$^{-1}$.

\textbf{Relation between resonance frequency and cavity size.} As previously mentioned, we do not have a proportionality relation between the resonance frequency and $1/L$ (see Fig.~\ref{fig2}.h). This discrepancy can be accounted for by considering the total phase acquired by the EMPs propagating beneath the input and output gates. This phase becomes non-negligible compared to that associated to the propagation of EMPs across the cavity when it becomes small enough, as it is the case in our experiment. From our numerical simulations, shown in dashed lines in Fig.~\ref{fig2}.h, the $v/L$ behavior is recovered for large cavities. However the increase of the frequency faster than $v/L$ for small cavities is well reproduced by our theoretical model.

\begin{figure}
    \centering
    \includegraphics[width = 0.5\textwidth]{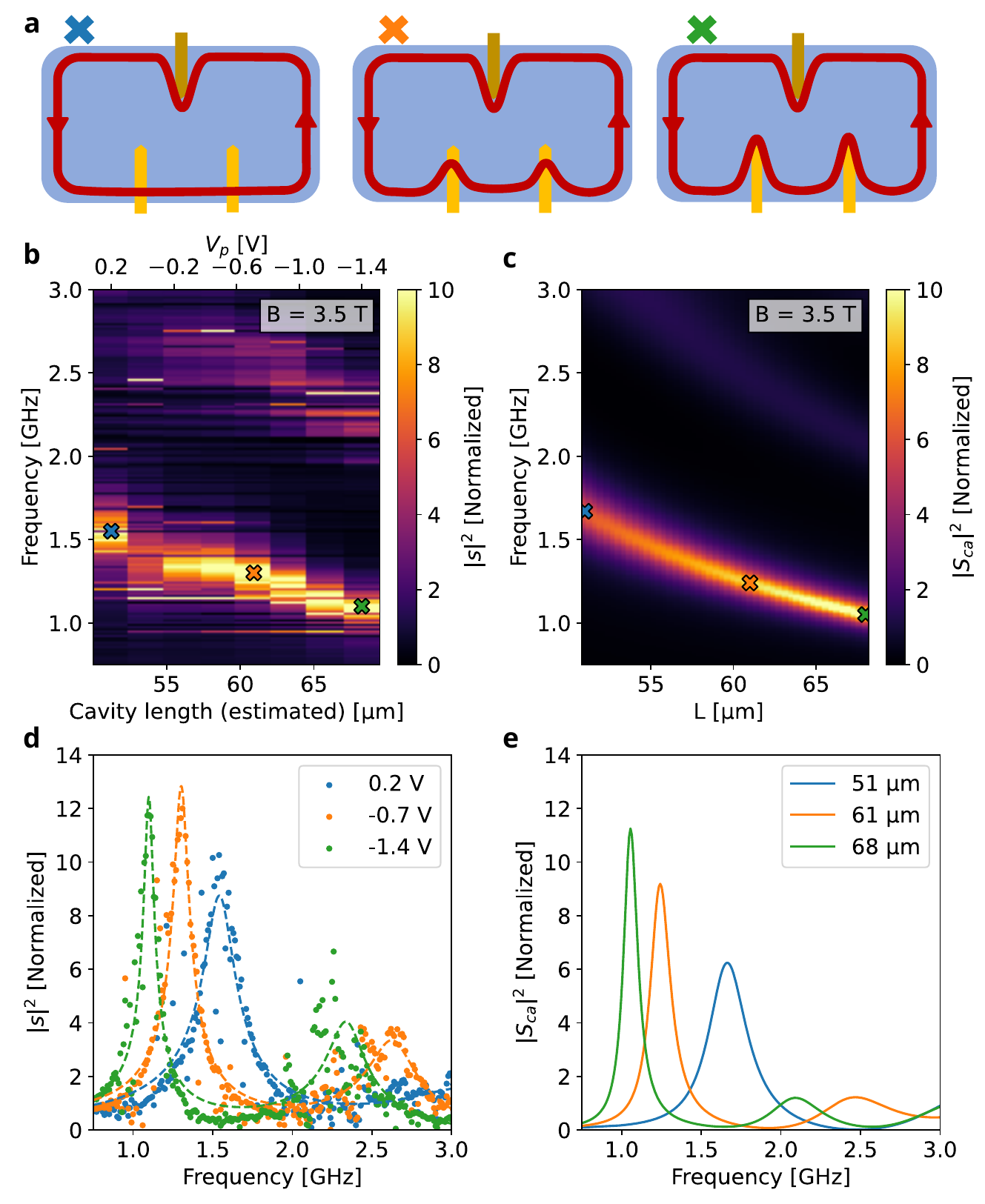}
    \caption{\textbf{Continuous control of the resonance frequency with electrostatic gates:} Using two  gates we can change continuously the perimeter of the structure and thus the frequency of the resonance. (a) At positive gate voltage the edge states go below the active gates (left). As the gate voltage is decreased, the edge states are moved towards the top of the gates (middle) and ultimately completely go around the gates (right). Doing so, the frequency of the resonant mode increases. (b) Frequency vs gate voltage map of the transmission amplitude of the sample. The markers correspond to the situations sketched in a. (c) Theoretical computation of frequency vs inverse of perimeter (d) Line cut of the transmission amplitude as a function of the frequency for 3 different values of the gate voltage as indicated in d. The dots represent the experimental data while the dashed lines correspond to a double Lorentzian fit of the data used to extract the quality factor $Q$. (e) Line cuts of the theoretical data shown in (c) taken at the same values as in d.}
    \label{fig4}
\end{figure}

In the simulations presented on Figs.~\ref{fig2}.d.-f., we observe the emergence of a higher frequency modes on top of the fundamental one for the larger island. The role of the FP's detection mirror is crucial when its dimension is comparable to that of the resonator. The gate then plays the role of filter. This can be mathematically observed through the sinc term that is present in the expression of the transmission and reflection terms $(r_j, t_j,t'_j)$. Physically, this simply accounts for the relation between the wavelength $\lambda = 2\pi/k$ of the EMP and the size of the detection gate. For instance, if we consider the charge density associated to the fundamental mode of a magnetoplasmon propagating on the edge of our system, the detection power of our probes will be related to the integrated field underneath those probes. With $\lambda$ the wavelength of the mode of interest, and considering a detection gate of size $\lambda/2$, we will be able to integrate a finite signal. However, the first harmonic of wavelength $\lambda/2$ and the associated signal integrated by the gate average to zero. This is the case for our smallest cavity shown in  Fig.~\ref{fig2}.c. In this case, the total size of the cavity is \SI{30}{\micro\meter} while the detection gate overlaps the EMPs over a length $L_c=\SI{11.3}{\micro\meter}\simeq \lambda/2$. For this reason, unlike previous works with large samples and gates of negligible size used for injection and detection \cite{talyanskii1992spectroscopy, talyanskii1994experimental, kumada2014resonant}, we are limited to the observation of only the fundamental and first harmonic of the cavity. This aspect is well reproduced in our theoretical modeling of the system.

Our theoretical calculation also anticipates higher order modes that are not observed experimentally. Other than the relative low amplitude of these modes, making them difficult to detect experimentally, their absence in the experimental data could be attributed to dipolar effects. These effects, linked to the non-circular geometry of the cavity, \cite{cano2013microwave, oblak2024} are not taken into account in our theoretical description.

\textbf{Continuous control of the cavity size.} The continuous control of the geometry of the cavity is essential in order to realize a microwave EMP interferometer in the future. Another way to control the size of the cavity is to use individual gates \cite{hashisaka2013} instead of completely closing QPCs as we did in Fig.~\ref{fig2}. Figure \ref{fig4} shows the result of this experiment. In this configuration the top gates are polarized to $V_G=\SI{50}{\milli\volt}$ (including the output gate, see methods section). The input gate is polarized to $V_{G}=\SI{-1.1}{\volt}$ leading to an edge following the contour of the gates instead of the chemically defined edge. This has the effect of changing the strength of the coupling of the resonant mode to the input gate. We then tune two gates at the bottom of the structure (see Fig.~\ref{fig4}.a). By applying a negative voltage bias on these two gates we can increase the size $L$ of the cavity from $\SI{54}{\micro\meter}$ (for $V_{G}=\SI{0.2}{\volt}$) to $\SI{71}{\micro\meter}$ (for $V_{G}=\SI{-1.4}{\volt}$). The result of this procedure is presented on Fig.~\ref{fig4}.b. As shown in Fig.~\ref{fig4}.d, we can extract the line shape of the resonance that clearly shows the shift of the resonance frequency as well as that of the first harmonic mode. Using a Lorentzian fit, we extract a quality factor $Q=f/\Delta f$ for the main resonant mode of $|s|^{2}$ that varies between $8$ (for $V_{G}=\SI{0.2}{\volt}$) and $18$ (for $V_{G}=-\SI{1.4}{\volt}$).

This experimental results can be well reproduced using the theory described previously as shown in Fig.~\ref{fig4}.c. In this figure, we plot the frequency as a function of the total length $L$ of the cavity instead of the QPC voltage as we cannot access the precise electrostatic of the system and the way the edge states are deformed by the applied voltage. Because this data was taken after thermally cycling the sample and for a different top-gate voltage compared to the data from Fig.~\ref{fig2}, in this case we extract a density $n_b=\SI{2e11}{\centi\meter^{-2}}$ and we have different fitting parameters $c_g=\SI{0.1}{\nano\farad\per\meter}$ and $a=\SI{1.8}{\micro\meter}$.

\section{Conclusion}

In conclusion, we have studied an electrostatically and magnetically tunable EMP resonator of very small size, such that the measurement probes are an integral part of the resonator and cannot be neglected in the theoretical description. We have shown that such resonator can be defined using electrostatic gates. 
Using top gates we are able to change the density of the 2DEG and screen the interaction between edge channels leading to a decrease of the velocity of the magnetoplasmons, a behavior well captured by our numerical simulations. We have compared our results to a theoretical description of our system and have found a good agreement between the two. In particular we find that the deviation from the expected $1/L$ trend of the EMP resonance frequency can be attributed to finite size effects of the input and output gates used in our experiment. We also explain the presence or absence of harmonics in our signal by frequency dependent filtering effects at the injection and detection gates. Finally, we have shown that we are able to change the size of the cavity at will by using fully or partially closed QPCs.

In future experiments, this electrostatic control should allow us to study resonances in a broad frequency range and, at the same time, reach surface areas that are small enough to control the magnetic field on the scale of the magnetic flux quantum. We will then be able to probe interference effects for elementary electronic or anyonic excitations within the resonator. This experimental platform constitutes the basis for future interferometric microwave experiments aimed at studying the statistics of quasiparticles in exotic phases of the quantum Hall effect. This platform also opens new possibilities to investigate the screening role of the mirrors on the cavity in extreme limits where the combined size of the mirror is larger than the free propagation length of EMPs. Moreover, it opens the way to a direct investigation of the spatial dependence of dissipation effects by probing the non-reciprocity of the capacitive coupling between an edge channel and a transmission line. Finally, it also enables studying non-linear plasmonic effects in resonators as well as parametric resonance processes through a dynamical control of the resonance frequency via a time dependent electrical control of the cavity's geometry.

\section*{Acknowledgments}
We thank E. Bocquillon, C. Dickel and B. Oblak for helpful discussions during the preparation of this manuscript. This project was funded in part by the "Agence nationale de la recherche" (ANR) project QuSig4QuSense (ANR-21-CE47-0012), AnyHall (ANR-21-CE30-0064). This project was also funded within the Quantera II Program that has received funding for the ElQURes project from the EU H2020 research and innovation program under the GA No 101017733, and with funding organization ANR (ANR-24-QUA2-0002). 

\section*{Author contributions}
U.G. and A.C. grew the substrates, Y.J. fabricated the samples. E.F., M.R., G.F. and G.M. designed and performed the experiment. E.F., M.R., G.F., B.P., E.B., J.-M.B.and G.M. and analyzed the data. G.R., H.S.-B., I.S. and P.D. developed the theoretical framework. E.F., G.R., and G.M. performed the numerical simulations. E.F., G.R. and G.M. wrote the manuscript with feedback from all the other authors.

\section*{Competing interests}

The authors declare no competing interests.

\section*{Data availability}
All data related to this paper are available on the zenodo platform under the following doi: 10.5281/zenodo.13644446. Additional raw data is available from the corresponding author upon reasonable request.

\section*{Code availability}
The code used to generate the figures in this paper is available on the zenodo platform under the following doi: 10.5281/zenodo.13644446. Additional raw data analysis code is available from the corresponding author upon reasonable request.

\section*{Methods}
\subsection{Base temperature}

All measurements presented here were performed in a dry dilution cryostat at a maximal temperature of \SI{30}{\milli\kelvin}. Other details about the methodology of this work, including data analysis and details of the theoretical modeling can be found in the supplementary information.

\subsection{Data normalization procedure}
\label{appendix/data-normalization}

Our experimental setup does not benefit from a low temperature calibration procedure that would allow us to extract the absolute value of the gain of our measurement chain. In order to overcome this difficulty and access a clear signal of the resonant modes present in our cavity we apply the following procedure to the raw signal amplified and collected.

Without such calibration, the gain of the whole rf setup and the background signal vary drastically over the \SI{8}{\giga\hertz} bandwidth of our measurement. However, at a fixed frequency, the variation of the magnetic field B only affects the sample, leaving the whole rf setup unchanged. Thanks to this property, we can numerically fix the gain of each frequencies and remove the background signal.

With $s_{\text{raw}}(f,B)$ the complex signal measured (amplitude and phase), the normalization procedure we use is the following :
\begin{itemize}
    \item First, at each frequency, we extract the average value  and the standard deviation of $s_\mathrm{raw}$ over $B$, 
    \begin{align*}
    \overline{s_\text{raw}}(f) &= \langle s_\text{raw}(f,B)\rangle_{B} \\ \sigma_{s_\text{raw}}(f) &= \sqrt{\langle|s_\text{raw}(f,B)-\overline{s_\text{raw}}(f)|^2\rangle_B}
    \end{align*}

    \item Then, we apply the transformation 
    \begin{align*}
    s(f,B) = \frac{s_\text{raw}(f,B)-\overline{s_\text{raw}}(f)}{\sigma_{s_\text{raw}}(f)}
    \end{align*}

\end{itemize}

Since we measure on a wide range of magnetic field $B$ and because the resonance width is small compared to this range, the mean value of  $\overline{s_\text{raw}}(f)$ is a good (and systematic) approximation of the background signal at $f$. $\sigma_{s_\text{raw}}$ is the average modulus of the background-less signal. Apart from the resonance, the signal measured is just a background noise, so $\sigma_{s_\text{raw}}$ is a good (and systematic) approximation of the background noise amplitude at $f$. Assuming this noise does not depend on the frequency $f$, $\sigma_{s_\text{raw}}$ is then a measure of the gain of our setup at a given frequency $f$.

This means that, in order to compute $\langle s_\mathrm{raw} \rangle_B$ and $\sigma^2_{s_\mathrm{raw}} $, we need the full $f,B$ map. For this reason, the data shown in figure 4 of the main text (b) are extracted from eight complete $f,B$ maps at various gate voltage $V_p$.

\subsection{Gate configurations}
\label{appendix/gate-configurations}

In order to optimize the detection of the resonant mode of the cavity, the electrostatic environment is different between the data shown in figure 2 and figure 4 of the main text. The values of the gate used to obtain figure 2 are presented in table \ref{gate_vals_fig2} while the gate configuration of figure 4 are presented in table \ref{gate_vals_fig4}. In those two tables the column \textit{Top gates} includes all gates except the QPCs or pads used to defined the limits of the cavity itself.

The densities are estimated from both the dc measurement of the Hall voltage and the rf maps.

\begin{table}[h]
    \centering
    \begin{tabular}{|c|c|c|c|}
    \hline
         $V_{G}^\mathrm{in}$ & $V_{G}^\mathrm{out}$ & Top gates & Density\\
        \hline        
          0~V &  0~V &  
          20~mV & $1.93\cdot10^{11}$~cm$^{-2}$\\
         \hline
    \end{tabular}
    \caption{Experimental gate configuration used in figure 2 of the main text.}
    \label{gate_vals_fig2}
\end{table}

\begin{table}[h]
    \centering
   \begin{tabular}{|c|c|c|c|}
    \hline
         $V_{G}^\mathrm{in}$ & $V_{G}^\mathrm{out}$ & Top gates & Density\\
        \hline        
          -1.1~V &  50~mV &  50~mV & $ 2\cdot10^{11}$~cm$^{-2}$\\
         \hline
    \end{tabular}
    \caption{Experimental gate configuration used in figure 4 of the main text.}
    \label{gate_vals_fig4}
\end{table}

\section{Characterization of QPCs}

\begin{figure}[h]
    \centering
    \includegraphics[width = 0.45\textwidth]{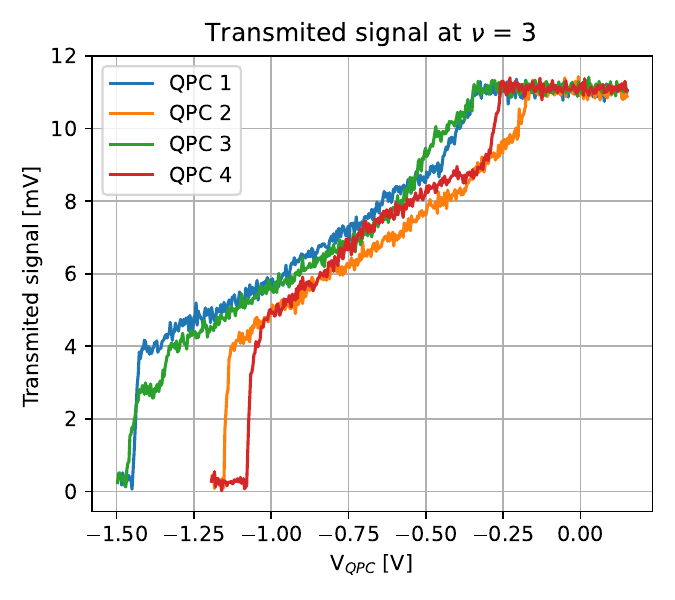}
    \caption{Transmitted signal as a function of the polarization of the QPCs. QPC~1 correspond to the QPC on the left of the sample, and QPC4 to the one on the right (see Fig 1 of the main text). The QPCs are considered to be closed when the transmitted signal reaches 0.}
    \label{figQPC}
\end{figure}

In order to change the shape and size of the cavities under study, we first characterize the behavior of QPCs in the dc regime. Figure \ref{figQPC} shows the voltage measured between contacts 1 and 2 as a function of the electrostatic voltage applied on the QPCs numbered from left to right following fig. 1 of the main text. The rf measurement were performed by closing some of the QPCs indicated in red in figs. 2.a.-c. of the main text using the voltage value for which the conduction falls to zero in the dc data.

\section{Characterization of the electronic density}
\label{appendix/dc-data}

\begin{figure}[h]
    \centering
    \includegraphics[width = 0.45\textwidth]{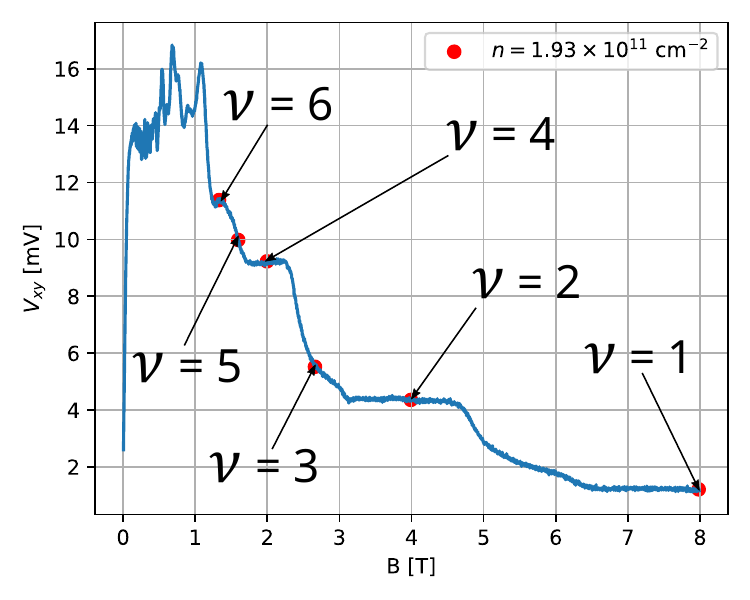}
    \caption{'Low frequency' measurement performed at 1.13kHz through Ohmic contacts. The data (blue lines) correspond to the transverse voltage drop of the 2DEG. The sample was measured in derivation with a 22~$\Omega$ resistor. The voltage is applied at contact 2, and the measurement is done at contact 1 (see figure 1 of the main text). The red dots represent the position of the Hall plateaus associated to an electronic density $n = 1.93\times10^{11}$cm$^{-2}$.}
    \label{figDC}
\end{figure}

The DC measurement allows us to identify the quantum Hall Plateaus as shown in figure \ref{figDC}. This data is obtained by applying a constant voltage drop on the sample (thanks to a 22~$\Omega$ resistor in derivation with contact 2 of figure 1 of the main text). The voltage is measured at contact 1. 

In this configuration, in dc measurement, all plateaus should appear at voltages increasing with the magnetic field. As we operate at finite frequency ($\approx 1$~kHz) this is not the case in our data. This can be easily explained by considering the role of the RC filters placed ahead of the sample. This makes the ensemble, sample plus capacitor, act as a low-pass filter. Sweeping the magnetic field changes the resistance of the sample and thus the cut-off frequency of the filter.

The sample presented in figure 1 of the main text has 4 large top gates that make it possible to tune the density of the 2DEG in the resonator. This change of density has consequences for both dc (in an open configuration) and rf (in a closed resonator configuration) measurements. By changing the density in the resonator, we can move the position of the plateaus. This will also displace the resonance frequency.

\section{Measurement procedures}
\label{appendix/measurement-procedures}

The data shown in figure 2 and in figure 4 of the main text were measured with two different setups. In this section, we will describe these two setups.

\begin{figure}[h]
    \centering
    \includegraphics[width = 0.45\textwidth]{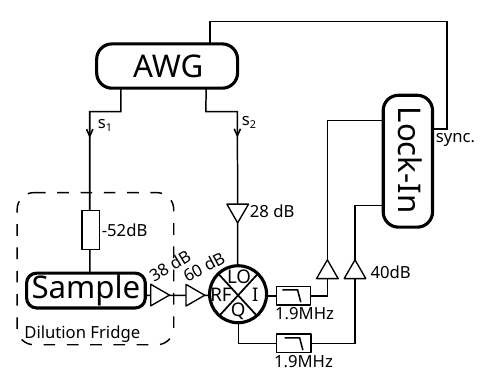}
    \caption{Schematic of the measurement setup used to get the data showed in figure 2 of the main text. The AWG generates two signals $s_1$ and $s_2$ at the selected frequency and with a fixed dephasing. $s_1$ is modulated by a square wave at 1~MHz. The IQ mixer extracts the two quadratures I and Q of the signal $s_1$ by multiplying $s_1$ and $s_2$. These two quadratures oscillate at the modulation frequency of $s_1$: 1~MHz. A HF2LI Lock-In measures these two quadratures.}
    \label{fighomodyne}
\end{figure}

The measurement setup depicted in figure \ref{fighomodyne} correspond to the one used to measure the data of figure 2 of the main text, it is a custom homodyne detection setup. We use an Arbitrary Wave Generator (AWG) to generate two signals $s_1$ and $s_2$ which are sine waves. $s_1$ is modulated by a square wave at 1~MHz and $s_2$ is a pure sine wave phase shifted compared to $s_1$. This dephasing is chosen so that the phase of $s_2$ and $s_1$ at the mixer are the same at $B=0$. By multiplying $s_1$ and $s_2$, the IQ mixer extracts the two quadratures of the signal coming out of the sample. These I and Q signals are oscillating at 1~MHz, the modulation frequency of $s_1$. They are measured using a Zurich-Instruments 50~MHz HF2LI Lock-In amplifier.

\begin{figure}[h]
    \centering
    \includegraphics[width = 0.35\textwidth]{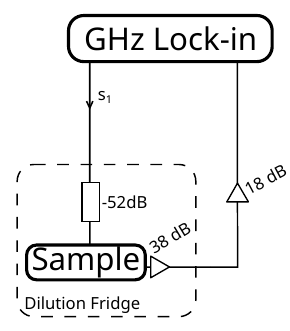}
    \caption{Schematic of the measurement setup used to obtain the data showed in figure 4 of the main text. In this configuration, the Zurich SHFLI GHz Lock-In generates the sine wave $s_1$ at the requested frequency and demodulates it after amplification at cryogenic and room temperature.}
    \label{figSHFLI}
\end{figure}

The measurement setup depicted in figure \ref{figSHFLI} corresponds to the one used to measure the data of figure 4 of the main text. For every gate point shown in Fig. 4 of the main text, a complete field/frequency map was acquired in order to perform the normalization procedure described in appendix A. Instead of a homemade GHz-lock-in detection, we use the 8.5~GHz Zurich-Instruments SHFLI Lock-in amplifier which is much faster and allows us to acquire the full maps necessary for normalization.

\section{Linearity of the resonator with input power}
\label{appendix/linearity}

\begin{figure}[h]
    \centering
    \includegraphics[width = 0.6\textwidth]{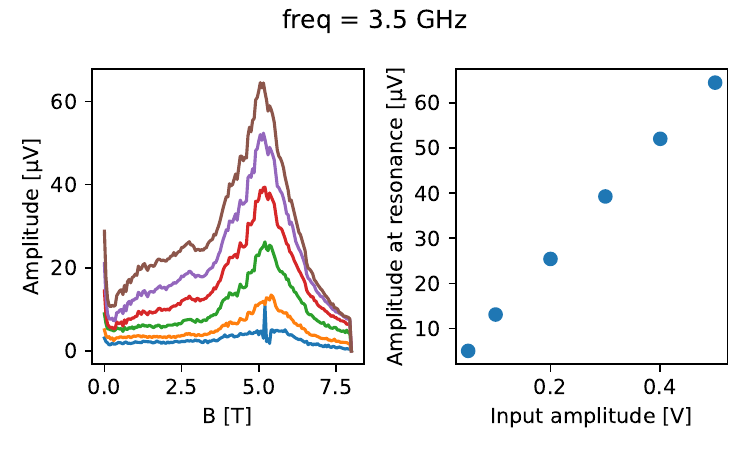}
    \caption{\textbf{Linearity of the resonator as a function of input voltage:} a. Field dependence of the lock-in signal for various values of the input rf voltage at $\SI{3.5}{\giga\hertz}$. b. Amplitude of the signal at resonance as a function of the input amplitude.}
    \label{figPowerLin}
\end{figure}

Figure \ref{figPowerLin} presents the evolution of the signal $|s|$ as a function of the amplitude of the source. This data was obtained from another sample, similar to the one used to obtain the data presented in the main text. From this data we observe the perfect linearity of our system.

\section{Magnetoplasmon velocity and dissipation}
\label{appendix/dissipation}

\begin{figure}[h]
    \centering
    \includegraphics[width = 0.6\textwidth]{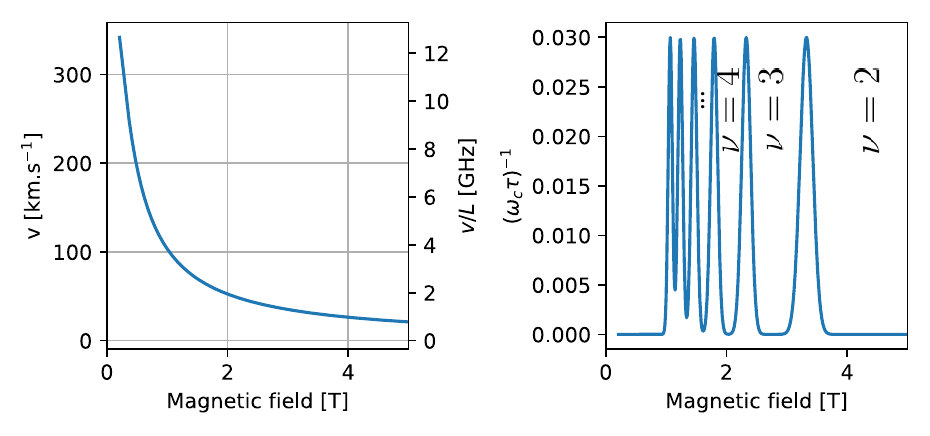}
    \caption{\textbf{Calculation of the velocity of magnetoplasmons:} (a) Estimation of the velocity of magnetoplasmons as a function of the magnetic field based on eq. \ref{speed_equation}. The right axis shows the associated resonance frequency of the bare resonator using the relation $f=v/L$ for a cavity perimeter of $\SI{27}{\micro\meter}$. (b) Dissipation of the system as a function of the magnetic field, following \cite{johnson2003} where a dissipation peak is added to a constant background in order to simulate the disappearance of the edge modes between the Hall plateaus.} 
    \label{figsup2}
\end{figure}

As we explain in the main text of this work, we use the description of the propagation of magnetoplasmons screened by a gate by Johnson {\it et al.} \cite{johnson2003}. In Fig.~\ref{figsup2}.a, we present the result of the calculation of the magnetoplasmon velocity 
\begin{equation}
v=\frac{a\gamma\omega_c}{1+\gamma r^2/4+(1/\omega_c^2\tau^2)}
\label{speed_equation}
\end{equation}
for a gate located $\SI{105}{\nano\meter}$ away from the 2DEG with a characteristic depletion size $a = \SI{2.8}{\micro\meter}$ and a bulk electronic density $n_b = \SI{1.93e11}{\centi\meter^{-2}}$. The same parameters are used in the calculation of Fig.~2 of the main text. The calculation leads to velocities of approximately $10^{5}$~m.s$^{-1}$ at $\SI{1}{\tesla}$ and $\SI{3e4}{\meter\per\second}$ at $\SI{3}{\tesla}$.

From these velocities we can use the naive formula for the frequency $f=v/L$ which leads to the data shown on the right axis of Fig.~\ref{figsup2}.a for a cavity length of $\SI{27}{\micro\meter}$. The resonance frequency of the cavity is found to be in the GHz range, just as in our experiment.

\begin{figure}[h]
    \centering
    \includegraphics[width = 0.6\textwidth]{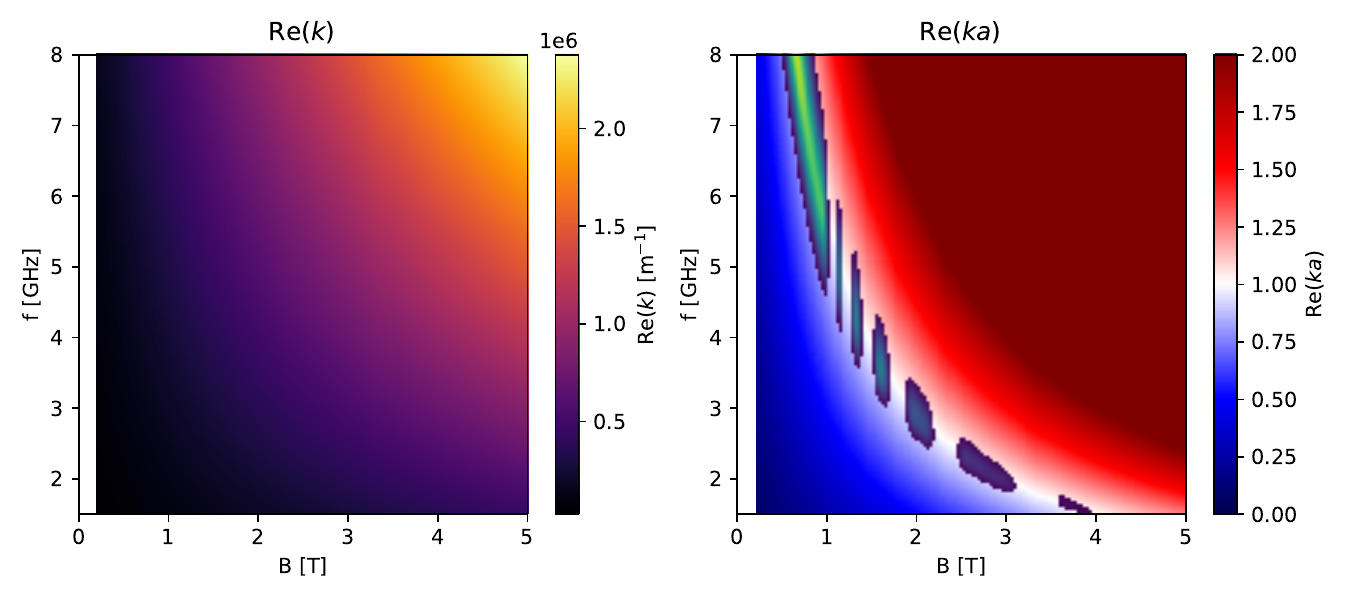}
    \caption{\textbf{Domain of validity of our simulations:} Computation of $ka$ as a function of $B$ and $f$. Resonance signal of the transmission amplitude for the small cavity has been superimposed over this map to indicate the regime in which we place ourselves.} 
    \label{figsup2_bis}
\end{figure}

To take into account the dissipation of magnetoplasmons in our system we introduce two effects. The first one relates to parasitic coupling to puddles and defects in the 2DEG that lead to a broadening of the resonant mode that we take into account by considering an imaginary part to $k$. On top of this dissipation during propagation, another source of dissipation is the disappearance of well defined edges states between the plateaus of the quantum Hall effect. Because EMPs are only well defined around integer values of the filling factors, there cannot be a resonance in-between those plateaus. Following \cite{johnson2003}, this is reflected in our calculation by considering an increase of the dissipation at field values equidistant from the center of two adjacent plateaus. This leads to a collection of peaks that form a continuum at low field. The model of dissipation $\omega_c\tau$ used in equation 1.a of the main text is presented on Fig.~\ref{figsup2}.b.

In Figure \ref{figsup2_bis}, we present the result of the calculation of the product $ka$ as a function of $f$ and $B$. In the model of \cite{johnson2003}, we should be in the limit of $ka\ll 1$. As can be seen here we actually reach $ka\simeq 1$ for the smaller cavity. We are thus placing ourselves at the limit of the approximations used by Johnson \& Vignale which might lead to over or under estimation of fitting parameters.

\section{Theoretical modeling}
\label{appendix/FP-scattering}

The theoretical modeling of the propagating signal between the input and output gates through the resonator is solved within an EMP scattering formalism. 

We model the first coupling region as an input transmission line with impedance $Z$ capacitively coupled to the resonator over the length $L_a$. The initial hypothesis of total electrostatic mutual influence translates into two conditions: the circuit neutrality and the capacitor neutrality. These mean that the charge stored in the metallic gate and in the transmission line are opposite and that the charge stored in the edge channel and in the gate are opposite.

\begin{figure}[h]
    \centering
    \includegraphics[width = 0.45\textwidth]{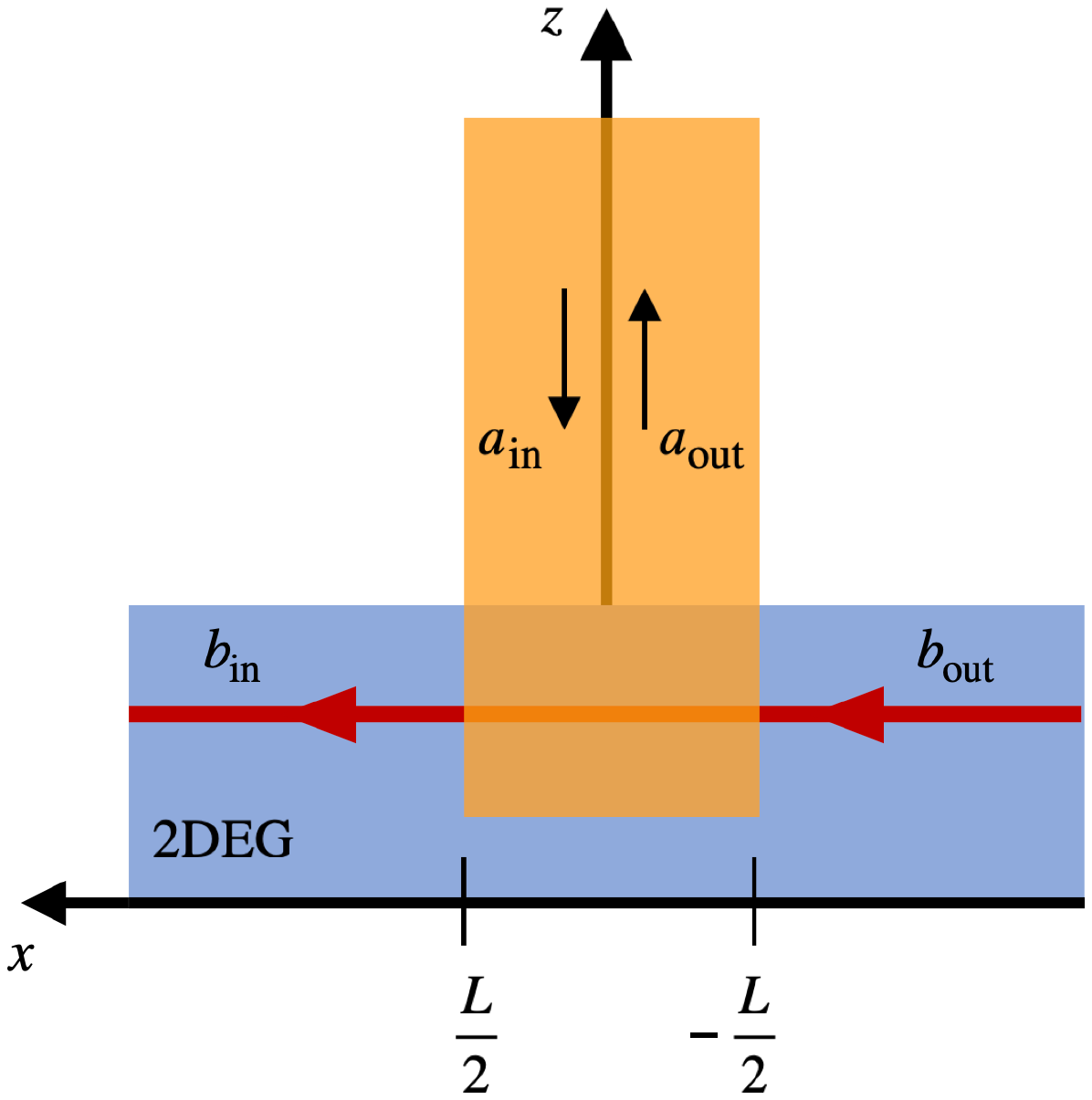}
    \caption{\textbf{Schematic view of the coupling region:} the yellow metallic gate, modeled as a transmission line with incoming and outgoing modes $a_{\mathrm{in}/\mathrm{out}}$, is capacitively coupled to the edge channel of the 2DEG over a region of length $L$.}
    \label{fig_mod_theo}
\end{figure}

In the coupling region, the equation of motion for the bosonic field $\phi_b (x,t)$ in the presence of an external potential $U_b(t)$ and the neutrality condition lead to
\begin{align}
&(\partial_t + v \partial_x) \phi_b(x,t)=\frac{e\sqrt{\pi}}{h}\, f(x)\,U_b(t) \,\, ,
\label{eom} \\
&Q_b(|x|\leq L/2,t)=Q(z=0,t)=-e\int_{-\frac{L}{2}}^{+\frac{L}{2}} n_b(x,t) \,dx    \label{charge} 
\end{align}
where $v$ is the EMP velocity along the channel, $n_b$ is the electronic density $n_b(x,t)=\partial_x \phi_b(x,t)/\sqrt{\pi}$ and $f(x)$ is the functional form of the considered square gate $f(x)=1$ if $|x|\leq L/2$ and $0$ otherwise. Then, the Fourier decomposition of the field $\phi_b (x,t)$ 
\begin{equation}
    \phi_b(x,t)=\int_0^{+\infty}\frac{d\omega}{2\pi}\,e^{i\left(\frac{x}{v}-t\right)}\,\phi_\omega(x)
    \label{phib}
\end{equation}
gives access to the relation
\begin{equation}
    \phi_\omega\left(\pm \frac{L}{2}\right)=-i\sqrt{\frac{\pi}{\omega}}\,b_{\mathrm{in}/\mathrm{out}}
    \label{bb}
\end{equation}
which links the field $\phi_b$ to the incoming or outgoing $b$-modes. 

Referring to Fig.~\ref{fig_mod_theo}, the transmission line is closed by the discrete capacitance $C_g$ at $z=0$ which appears as a boundary condition. The potential drop across the capacitance can be written as
\begin{equation}
U_b(t)-V_g(t)=\frac{Q(0,t)}{C_g}
\label{pot}
\end{equation} 
where $U_b$ is the potential felt by the electron propagating in the edge channel of the 2DEG and 
$V_g$ is the potential on the capacitor plate connected to the transmission line. The potential $U_b(t)$ is found by knowing the Fourier decomposition of the total charge $Q(0,t)=Q_\mathrm{in}(0,t)+Q_{\mathrm{out}}(0,t)$ \begin{equation}
    Q(0,t)=e \sqrt{\frac{R_K}{2Z}}\int_0^{+\infty}\frac{d\omega}{2\pi}\frac{1}{\sqrt{\omega}}\big[\big(a_{\mathrm{in}}+a_{\mathrm{out}}\big)+\mathrm{h.c.}\big]e^{-i\omega t}
    \label{q0}
\end{equation}
and of the drop potential 
\begin{equation}
V_g(t)=-Z \left(\partial_t Q(0,t)\right).
\label{vg}
\end{equation}

As explained in the main text, to understand the experimental data, we take into account energy dissipation for the EMPs into the phonon bath. This is phenomenologically introduced through a quadratic contribution proportional to $\omega^2$ in $\mathrm{Im}(k(\omega))$. This quadratic dependence is chosen to be compatible with a discrete circuit elements description at low frequency~\cite{hashisaka2013, bocquillon2013, rodriguez2020, rebora2021}. Then, equation~\eqref{eom} in the frequency space is rewritten as
\begin{equation}
\big[-ik(\omega)+\partial_x\big]\tilde{\phi}_b(x,\omega)=\frac{e\sqrt{\pi}}{vh}f(x) \tilde{U}_b(\omega)
\label{eomdiss}
\end{equation}

Through Eqs.~\eqref{q0} and~\eqref{vg} we find the expression in the frequency space for $\tilde{U}_b(\omega)$. This, together with Eqs.~\eqref{phib} and~\eqref{bb} for the bosonic field, allows to solve the equation of motion in~\eqref{eomdiss}. Combining it together with Eq.~\eqref{charge}, we finally find the scattering matrix for the plasmonic modes $a$ and $b$ as
\begin{equation}
    \begin{pmatrix}
        a_{\mathrm{out}} \\ b_{\mathrm{in}}
    \end{pmatrix}=\begin{pmatrix}
        r_a & t_a \\
        t'_a & r'_a
    \end{pmatrix}\begin{pmatrix}
         a_{\mathrm{in}} \\ b_{\mathrm{out}}
    \end{pmatrix}.
    \label{scatta}
\end{equation}
Through Eqs.~4 and~5 of the main text we express the transmission from the line to the edge channel ($t'_a$) and from the edge channel to the line ($t_a$) while the coefficient $r'_a$ in Eq.~6 of the main text refers to the reflection in the edge channel. Then, for the last coefficient $r_a$ we have
\begin{equation}
    r_a=-e^{-\mathrm{Im}(k(\omega))\frac{L_a}{2}}\left[1+ e^{ik(\omega)\frac{L_a}{2}}\alpha_a \mathrm{sinc}\left(k(\omega)\frac{L_a}{2}\right)\times \left(1+\mathrm{Re}(k(\omega))\frac{\mathcal{R}}{2}\frac{L_a}{\alpha_a}\right)\right]D_a^{-1}.
\end{equation}

It should be noted that the introduction of dissipation leads to a breakdown in the reciprocity of the scattering matrix in Eq. \eqref{scatta}, i.e. $t_a\neq t'_a$. This break would allow us to have a direct observation on the role of dissipation by studying the coupling region between the transmission line and the edge channel.

The scattering process at the level of the second coupling region between the resonator and the output transmission line, with impedance $Z$, is considered in the same spirit as before. Here, the $b$ mode has acquired (lost) a phase $X_b$ when traveling toward (from) the metallic gate. We can then write
\begin{equation}
    \begin{pmatrix}
        c_{\mathrm{out}} \\ b_{\mathrm{out}}e^{-iX_b}
    \end{pmatrix}=\begin{pmatrix}
        r_c & t_c \\
        t'_c & r'_c
    \end{pmatrix}\begin{pmatrix}
         c_{\mathrm{in}} \\ b_{\mathrm{in}}e^{iX_b}
    \end{pmatrix}
    \label{scattc}
\end{equation}
where the new coefficients assume the same functional form like the previous one with the difference in length $L_c$ and capacity $C_{g,c}$. 

Our interest in studying the resonant mode, in transmission or reflection, leads us to write 
\begin{equation}
    \begin{pmatrix}
        c_{\mathrm{out}} \\ a_{\mathrm{out}}  \end{pmatrix}=S(\omega)\begin{pmatrix}
         c_{\mathrm{in}} \\ a_{\mathrm{in}}
    \end{pmatrix}
\end{equation}
which relates the transmitted mode $c_{\mathrm{out}}$ and the reflected one $a_{\mathrm{out}}$ in terms of the incoming $a_{\mathrm{in}}$ and $c_{\mathrm{in}}$. By combining Eqs.~\eqref{scatta} and~\eqref{scattc} and solving in terms of $a$ and $c$ modes we obtain the final expression for the scattering matrix

\begin{widetext}
\begin{equation}
S(\omega)=\begin{pmatrix}
    S_{cc} & S_{ca} \\
    S_{ac} & S_{aa}
\end{pmatrix}=\frac{1}{1-r'_a r'_c\,e^{2iX_b}}
    \begin{pmatrix}
    r_c-r'_a\,(r_a r'_a -t_a t'_a)\,e^{2iX_b} & t'_a t_c e^{iX_b} \\
    t_a t'_c e^{iX_b} & r_a-r'_c\, (r_a r'_a -t_a t'_a)\,e^{2iX_b}
    \end{pmatrix}.
    \label{fullscatt}
\end{equation}
\end{widetext}
Notice that in the low frequency limit the scattering matrix reduces to the identity matrix and in absence of dissipation the coefficients satisfy the unitarity conditions $|S_{ca}|^2+|S_{cc}|^2=1$ which ensure energy conservation within the system. Finally, the scattering matrix in Eq.~\eqref{fullscatt} allows to look at the resonances of the considered cavity by studying the transmission amplitude ($S_{ca}$) or by focusing on the reflection one ($S_{cc}$) as in~\cite{mahoney2017chip}. 

\section{Role of the capacitance}

\begin{figure}[h!]
    \centering
    \includegraphics[width = 0.8\textwidth]{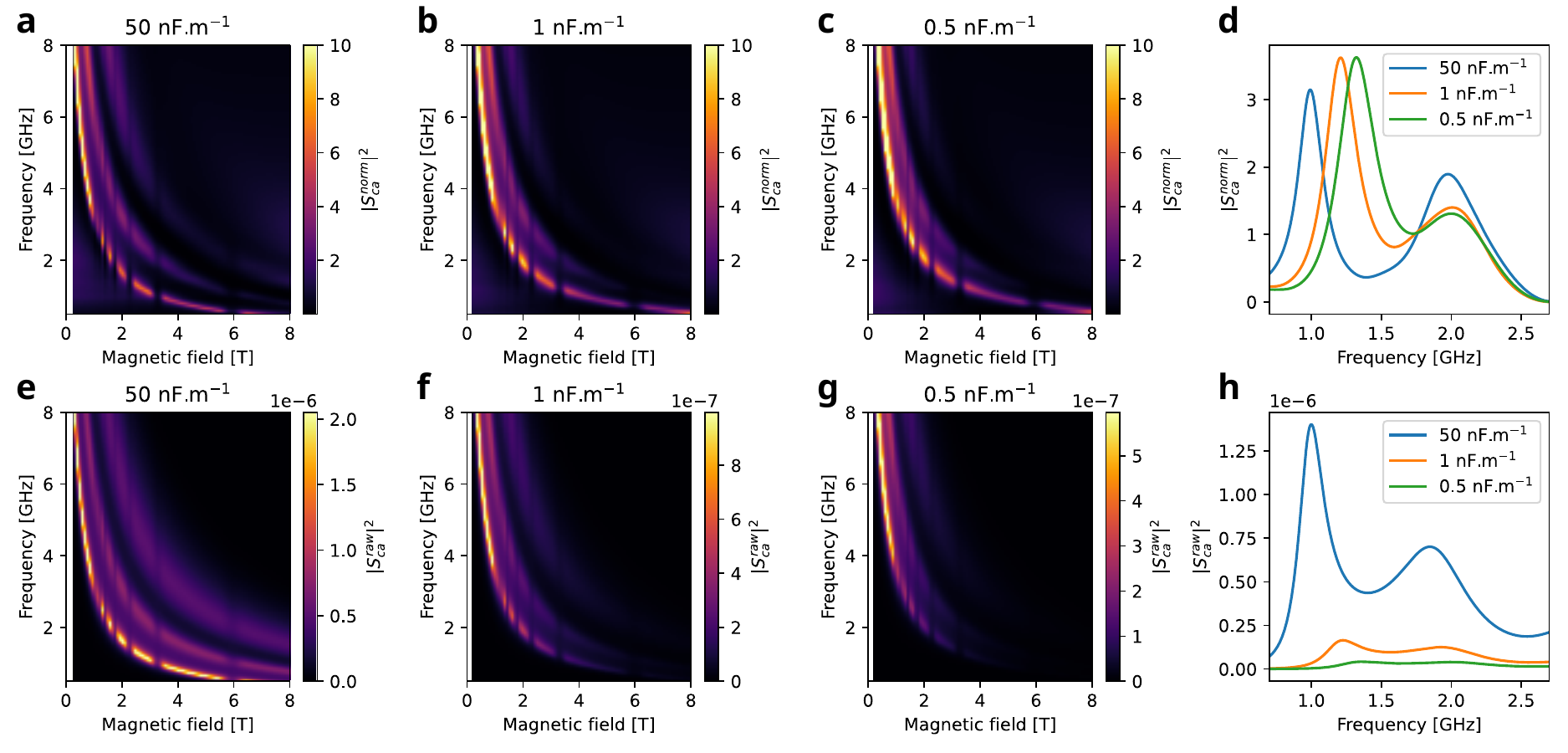}
    \caption{\textbf{Gate effect on transmission signal:} (a)-(c) Normalized transmission signal $|S_{ca}|^2$ calculated for three different value (50, 1 and 0.5 nF.m$^{-1}$) of the lineic capacitance $C_g^{(l)}$. (d) Cuts of maps a-c taken at 3.5~T showing the line shape of the resonance as a function of frequency for the same value of the capacitance. (e)-(g) Raw transmission maps and (h) raw cuts without applying the normalization procedure.}
    \label{figsup1}
\end{figure}

The capacitance between the EMPs and the input or output gates $C_{g,j}$ appears in the calculation of the elements of the transmission matrix $S$ and in particular in the quantity $S_{ca}$ that relates the incoming and the outgoing amplitudes of the field (see appendix F for details). This capacitance guarantees that the edge modes connect to the outside world to ensure a measurable intensity of the experimental signal. In figure \ref{figsup1} we present the result of the calculation of $|S_{ca}|^2$ for three difference values of the lineic capacitance (50, 1 and 0.5 nF.m$^{-1}$) for both the raw signal (e to g) and the normalized data (a to c). From these calculation we observe that as the capacitance increases, so does the amplitude of the measured signal. However, increasing the capacitance has also the effect of shifting the resonances. This can be seen in the cuts taken at 3.5~T shown in figures \ref{figsup1}.d and h that show resonance evolving from 1~GHz to 1.4~GHz.

\section*{References}
\bibliography{biblio}

\end{document}